\documentclass{sig-alternate}
\usepackage[
  pass,
]{geometry}
\usepackage{graphicx} 
\usepackage{subfigure} 
\usepackage[utf8]{inputenc}

\allowdisplaybreaks
\usepackage{algorithm}
\usepackage{algorithmic}
\usepackage{amssymb}
\usepackage{amsmath}
\usepackage{bm}
\usepackage{mathtools}
\usepackage{paralist}

\usepackage{etoolbox}
\newtoggle{ARXIV}
\toggletrue{ARXIV}

\usepackage{booktabs}
\usepackage[pdfpagelabels=false]{hyperref}
\usepackage[usenames,dvipsnames]{xcolor}
\usepackage{pgf}

\newdef{proposition}{Proposition}

\DeclareMathOperator{\rel}{rel}
\DeclareMathOperator{\rank}{rank}

\DeclareMathOperator{\sign}{sign}

\DeclareMathOperator{\Prob}{P}
\newcommand{\X}{{\mathcal{X}}}           
\newcommand{\x}{\bm{x}}                  
\newcommand{\xsample}{\bm{X}}       
\newcommand{\y}{\bm{y}}                  
\newcommand{\ypart}{y}                   
\newcommand{\Util}{U}                    
\newcommand{\Utilhat}{\hat{\Util}}       
\newcommand{\util}{u}                    
\newcommand{\len}{|\xsample|}                    
\newcommand{\utilapx}{\tilde{\util}}     
\newcommand{\utilopt}{\util^*}           
\newcommand{\Diff}{\Delta}                    
\newcommand{\Diffhat}{\hat{\Diff}}       
\newcommand{\weight}[2]{\lambda(#1\mid#2)}               

\newcommand{\E}{{\mathbb{E}}}            
\newcommand{\systems}{{\mathcal{S}}}     
\newcommand{\system}{{S}}                
\newcommand{\user}{{a}}                  
\newcommand{\n}{{n}}                     
\newcommand{\data}{\mathcal{D}}          
\newcommand{\Ind}{\mathbb{I}}            

\begin{document}
\setcopyright{acmcopyright}

\doi{10.475/123_4}

\isbn{123-4567-24-567/08/06}

\acmPrice{\$15.00}
\conferenceinfo{ICTIR}{'16 Delaware USA}

\title{Unbiased Comparative Evaluation of Ranking Functions}
\iftoggle{ARXIV}{
\numberofauthors{4} 
\author{
\alignauthor
Tobias Schnabel\\
       \affaddr{Cornell University, Ithaca, NY}\\
       \email{tbs49@cornell.edu}
\alignauthor
Adith Swaminathan\\
       \affaddr{Cornell University, Ithaca, NY}\\
       \email{adith@cs.cornell.edu}
\and
\alignauthor Peter I. Frazier\\
       \affaddr{Cornell University, Ithaca, NY}\\
       \email{pf98@cornell.edu}
\alignauthor Thorsten Joachims\\
       \affaddr{Cornell University, Ithaca, NY}\\
       \email{tj@cs.cornell.edu}
}
}{
\numberofauthors{1} 
\author{
\alignauthor
<omitted>
       \affaddr{\phantom{Address}}
       \email{\phantom{author@institution.com}}
}
}

\maketitle
\begin{abstract}
Eliciting relevance judgments for ranking evaluation is labor-intensive and costly, motivating careful selection of which documents to judge. Unlike traditional approaches that make this selection deterministically, probabilistic sampling has shown intriguing promise since it enables the design of estimators that are provably unbiased even when reusing data with missing judgments. In this paper, we first unify and extend these sampling approaches by viewing the evaluation problem as a Monte Carlo estimation task that applies to a large number of common IR metrics.
Drawing on the theoretical clarity that this view offers, we tackle three practical evaluation scenarios: comparing two systems, comparing $k$ systems against a baseline, and ranking $k$ systems.
For each scenario, we derive an estimator and a variance-optimizing sampling distribution while retaining the strengths of sampling-based evaluation, including unbiasedness, reusability despite missing data, and ease of use in practice. In addition to the theoretical contribution, we empirically evaluate our methods against previously used sampling heuristics and find that they generally cut the number of required relevance judgments at least in half.
\end{abstract}

\category{H.4}{Information Systems Applications}{Miscellaneous}

\terms{Algorithms, Design, Performance, Measurement}

\keywords{Importance Sampling, Evaluation, Crowd Sourcing, Pooling}

\section{Introduction}
\label{sec:intro}
Offline evaluation of retrieval systems requires annotated test collections that take substantial effort and cost to amass.
The most significant cost lies in eliciting relevance judgments for query-document pairs, and
the size of realistic test collections makes it infeasible to annotate every query-document pair in the corpus.
This has spurred research on intelligent choice of pairs to judge. 
Analogous annotation problems also exist in other domains, like machine translation or sequence tagging in natural language processing. Moreover, with the advent of crowd-sourced annotations for applications like image recognition and protein sequencing, the problem of judgment elicitation has become even more relevant.

Unfortunately, if the whole corpus is not judged, the missing judgments may bias the performance estimates. There are two broad approaches to addressing this problem.
The first develops new evaluation measures that are robust to incompletely judged test collections \cite{Sakai2008}. 
For such measures, heuristics like pooling can then be employed effectively, but the evaluation measure may not precisely capture the notion of quality one requires.
The other approach is to leave the design of the evaluation measure as unrestricted as possible, but instead design general evaluation methodologies and estimators that guarantee unbiased estimates 
even under missing data \cite{aslam2006statistical,Yilmaz2006,Pavlu2007,yilmaz2008simple}. We follow this second approach, specifically focusing on sampling approaches that possess the following desirable properties:
\begin{enumerate}
\item Unbiasedness. On average, estimates have no systematic error, and behave like we had judged all query-document pairs.
\item Reusability. Collecting judgments is labor-intensive and costly, and sampling-based approaches allow reuse of past data without introducing bias.
\item Statelessness. We often need to collect tens of thousands of judgments. Sampling is embarrassingly parallel and can be done in a single batch.
\item Sample Efficiency. Sampling distributions can be designed to optimize the number of judgments needed to confidently and accurately estimate a metric.
\end{enumerate}

In this paper, we focus on three comparative evaluation scenarios that frequently arise in practice, and derive new estimators and principled sampling strategies that substantially improve sample efficiency while retaining the other desirable properties of sampling-based evaluation. 
In particular, we investigate the problem of estimating the performance difference between two systems, the problem of estimating $k$ systems' performance relative to a baseline, and the problem of estimating a ranking of $k$ systems.
For all three scenarios, we propose importance-weighted estimators and their variance-optimizing sampling distributions, enabling the estimators to elicit relative performance differences much more efficiently than previous sampling approaches.
We show that these estimators apply to any linearly decomposable performance metric (e.g., DCG, Precision@k), and that they are unbiased even with missing judgments. 
In addition to these theoretical arguments, empirical results show that our estimators and samplers can substantially reduce the number of required judgments compared to previously used sampling heuristics, making them a practical and effective alternative to pooling and heuristic sampling methods. Beyond these specific contributions, the paper more generally contributes a unified treatment of all prior sampling approaches in terms of Monte Carlo estimation and its theory, providing a rigorous basis for future research.

\section{Related Work}
\label{sec:related}
The Cranfield methodology \cite{Cleverdon1991} using pooling \cite{SparckJones1975} is the most established approach to IR evaluation.
Pooling aims to be fair to all submitted systems by exhaustively judging the top-ranked document from all systems, hoping for good coverage of all relevant documents for a query (implicitly through diversity in the submitted runs).
However, pooling bias is a well known problem when re-using these pooled collections to evaluate novel systems that retrieve relevant but unjudged documents \cite{Zobel1998} -- see \cite{Sanderson2010} and the references therein for a detailed overview.
Attempts have been made to correct this pooling bias, either using a small set of exhaustively judged queries \cite{Webber2009} or using sample corrections that apply for MAE and Precision@k \cite{Buttcher2007,Lipani2016}.

Generally, however, the size of today's corpora has prevented complete judging of test collections, and this 
has driven research on evaluation strategies that are robust to missing judgments \cite{Carterette2009}.
One approach to handling incomplete judgments is to define an IR metric that is robust to missing judgments, like Bpref \cite{Buckley2004}, RankEff \cite{Ahlgren2006} and Rank-Biased Precision \cite{Moffat2008}. Sakai and Kando \cite{Sakai2008} provide an excellent review of these approaches.  
Another approach, and the one we build on in this paper, uses random sampling of ranked documents to construct a collection of judgments \cite{Cormack2006,aslam2006statistical,Yilmaz2006,Pavlu2007,yilmaz2008simple}. We unify all these sampling approaches by viewing them as Monte Carlo estimates of IR metrics and extend them to relative comparisons.

The idea of relative comparisons rather than absolute evaluation of IR measures has been studied before.
Deterministic elicitation schemes have been proposed for differentiating two systems based on AP \cite{carterette2006minimal},
and to rank multiple systems according to Rank-Biased Precision \cite{Moffat2007}.
More recently, multi-armed bandit approaches have been studied to construct judgment pools \cite{Davide2016}.
These schemes suffer from the same bias that plagues pooling when comparing new systems.
We extend the provably unbiased sampling approaches to these comparative scenarios, inheriting the improved sample efficiency of relative comparisons while yielding re-usable test collections.
We anticipate future work that combines the simplicity of batch sampling with the sample efficiency of active learning and bandit algorithms to adaptively elicit judgments.

We note that the sampling approach we take here naturally incorporates noisy judgments, making it suitable for many tasks involving crowd-sourcing. Existing works have heuristically resolved noisy judgments as a pre-processing step before constructing the test collection \cite{hosseini2012aggregate,ye2013combining}.


Ideas from Monte Carlo estimation and importance sampling \cite{mcbook} have been successfully applied in closely related problems like unbiased recommender evaluation \cite{Schnabel2016,li2011unbiased}, although in those applications, the sampling distribution is typically not under the experimenter's control.
Finally, a related problem to the judgment elicitation problem we study here is that of picking the most informative set of queries \cite{Guiver2009}.
Our Monte Carlo formulation can offer a reasonable starting point to answer this question as well.

\section{Sampling Based Evaluation}

To support our novel sampling approaches to comparative evaluation, described in the following three sections, we first lay out a unified framework of Monte Carlo estimation for sampling-based evaluation of a single system, unifying existing IR work \cite{aslam2006statistical,Yilmaz2006,Pavlu2007,yilmaz2008simple} with the extensive literature and theory in Monte Carlo estimation. 

\subsection{Illustrative Example}

Consider a retrieval system $\system(\x)$ that maps each input query $\x$ to a ranking $\y$. Given a set of $\len$ queries $\xsample$, we would like to estimate the average Discounted Cumulative Gain (DCG) with depth cut-off 100 of $\system$ on $\xsample$,
\vspace{-0.8em}
\begin{eqnarray}
  DCG@100(\system)  & = & \frac{1}{\len}\sum_{\x \in \X} \sum_{r=1}^{100} \frac{\rel(\x,\system(\x)_r)}{\log(1+r)} \nonumber \\
                & = & \sum_{(\x,r)} \frac{\rel(\x,\system(\x)_r)}{\len \cdot \log(1+r)} \nonumber \\
                & = & \sum_{(\x,r)} V(\x,r) \nonumber .
\end{eqnarray}
$\system(\x)_r$ is the document at rank $r$ in the ranking of $\system$ for query $\x$, and $\rel(\cdot)$ denotes its assessed relevance. The key insight behind sampling-based evaluation is the following: we do not need to know all summands $V(\x,r)$ in this large sum to get a good estimate of $DCG@100(\system)$. In particular, even if we just uniformly at random sample $\n$ query-document pairs $D=((\x_1,r_1),...,(\x_n,r_n))$ and elicit the value of $V(\x,r)$ for those, the following average is a reasonable estimate of $DCG@100(\system)$ even when $\n < 100 \cdot \len$.
\begin{equation}
   DCG@100(\system) \approx \frac{100 \cdot \len}{\n} \sum_{(\x_i,r_i) \in D} V(\x_i,r_i) \nonumber.
\end{equation}
But what are the quality guarantees we can give for such estimates? Is uniform sampling the best we can do? What about other performance measures? And what about statistical testing and comparisons of multiple systems? To address these questions, we now formalize sampling-based evaluation in the framework of Monte Carlo estimation.

\subsection{Formalizing Evaluation}

A ranking system $\system(\x)$ (e.g. search engine, recommendation system) maps an input $\x \in \X$ (e.g. query, user context) to a ranking $\y$. Each predicted ranking $\y=\system(\x)$ has a certain utility $\Util(\x,\y)$ for a given $\x$ which quantifies the quality of ranking $\y$ for $\x$. To aggregate quality over multiple $(\x,\y)$ pairs, virtually all evaluation approaches use the expected utility over the distribution $\Prob(\x)$ as a summary of the overall quality of a system.
\begin{equation*}
\Util(\system)=\mathbb{E}_{\Prob(\x)}[\Util(\x, \y)] = \int \Util(\x,\y) d\Prob(\x), 
\end{equation*}
where, again, $\y = \system(\x)$ refers to the output of $S$ given $\x$.
Most often though, we wish to evaluate \emph{multiple} systems, i.e.,
we have a number of systems $\systems = \{\system_1, \ldots \system_k \}$ which we wish to evaluate on a distribution of inputs $\Prob(\x)$. 
Taking web-search as an example, $\systems$ would be a collection of retrieval systems, $\Prob(\x)$ the distribution of queries, and the utility of a ranking $\Util(\x,\y)$ would be measured by some IR metric like Precision@$10$. We will now discuss how to obtain values of $\Util(\x, \y)$.


\subsection{Linearly Decomposable Metrics}
Since assessing the utility $\Util(\x,\y)$ of a complete ranking $\y$ is difficult for human assessors, the utility of $\y$ is typically aggregated from the utilities of the individual documents it contains. 
Formally, we assume $\Util$ decomposes as a sum of the utilities of the parts
\vspace{-0.3em}
\begin{equation}
\Util(\x, \y) = \sum_{\ypart \in \y} \weight{\ypart}{\y} \, \util(\x, \ypart).
\label{eq:lindecomp}
\end{equation}
The weights $\weight{\ypart}{\y} \ge 0$ with which the utility of each document $\ypart$ of ranking $\y$ enters the overall utility is defined by the particular performance measure. The utilities $\util(\x, \ypart) \in \mathbb{R}^{+}$ refer to the individual utilities of each part $\ypart$ in $\y$.
Taking Precision@k as an example, $\util(\x, \ypart) = \rel(\x, \ypart) \in \{0, 1\}$ denotes binary relevance of document $\y$ for query $\x$, and the weights $\weight{\ypart}{\y}$ are $1/k$ if $\ypart$ appears among the top $k$ documents in the ranking $\y$, and zero otherwise.

The top half of Table~\ref{tbl:itembased} shows more examples of performance measures that are linear functions of the individual parts $\ypart$ of $\y$. The bottom half of Table~\ref{tbl:itembased} presents examples of performance measures whose natural decomposition of $\y$ is into pairs of variables, i.e.,
$y = (\tilde{\ypart}_1, \tilde{\ypart}_2)$. 
These examples demonstrate the wide applicability of the decomposition in Eq.~\eqref{eq:lindecomp}. Furthermore, one can estimate normalized measures (e.g. AP, NDCG) by taking ratios of estimated $\Util$ at the expense of a typically small bias \cite{aslam2006statistical,yilmaz2008simple}. 
Other structured prediction tasks, like sequence labeling, parsing, and network prediction, use similar part-based performance measures, and much of what we discuss can be extended to evaluation problems where $\y$ is a more general structured object (e.g. sequence, parse tree).

\begin{table}[tb]
  \centering
    \begin{tabular}{lll}
    \toprule
    Metric & $\util(\x, \ypart)$ & $\weight{\ypart}{\y}$ \\
    \midrule
    $Prec@k$ & $\rel(\x, \ypart) \in \{0, 1\}$   & $\mathbf{1}_{\rank(\ypart) \leq k}/k$ \\
    $DCG$ & $\rel(\x, \ypart) \in [0, M]$  & $ 1\!/\!\log\!\left( 1 + \rank \left(\ypart\right)\right)$ \\
    $Gain@k$ & $\text{rel}(\x, \ypart) \in [0, M]$ & $\mathbf{1}_{\rank(\ypart) \leq k}/k$ \\
    $MAE$ & $|\text{error}(\x, \ypart)| \in [0, M]$ & $1/|\y|$ \\
    $MSE$ & $(\text{error}(\x, \ypart))^2 \in [0, M]$ & $1/|\y|$ \\
    \emph{RBP-p} \cite{Moffat2008} & $\rel(\x, \ypart) \in \{0, 1\}$ & $(1-p)/p^{\rank(y)}$ \\
    \midrule
    $\# of Swaps$ & $\text{swapped}(\x, \ypart) \in \{0, 1\}$   & $1/|\y|$ \\
    $wSwaps$ & $\text{swapped}(\x, \ypart) \in \{0, 1\}$   & $1/(|\y|\cdot \rank(\tilde{y_1})$ \\
     &    &  $\cdot \rank(\tilde{y_2}))$ \\
    $AP$ & $\rel(\x, \tilde{y}_1)\cdot\rel(\x, \tilde{y}_2)$ & $\mathbf{1}_{\rank(\tilde{y_2}) \leq \rank(\tilde{y}_1)}$ \\
    &  $\in \{0, 1\}$   & $/ \left( R \cdot \rank(\tilde{y}_1) \right)$ \\
    \bottomrule
    \end{tabular}%
  \caption{A selection of popular metrics that can be written as expectations over single item judgments (top) or pairwise judgments (bottom). $R = \sum_{\tilde{y}} \rel(x, \tilde{y})$.}
  \vspace{-0.8em}
  \label{tbl:itembased}%
\end{table}%


\subsection{Evaluation as Monte Carlo Estimation}

Previous work has realized that one can use sampling over the documents $\ypart$ in the rankings $\y=\system(\x)$ to estimate Average Precision and NDCG \cite{aslam2006statistical,Yilmaz2006,Pavlu2007}. 
However, the idea of sampling over the components $\ypart$ applies not only to these performance measures, but to any linearly decomposable performance measure that can be written in the form of Eq.~\eqref{eq:lindecomp}. Making the connection to Monte Carlo methods, we start by defining the following distribution over the documents $\ypart$ of a ranking $\y=\system(\x)$,
\begin{align*}
\Pr(\ypart \mid \x; \system) &= \frac{\weight{\ypart}{\system(\x)}}{\sum_{\ypart'} \weight{\ypart'}{\system(\x)}}.
\end{align*}
To simplify the exposition, we assume that the weights are scaled to sum to $1$ (i.e., $\sum_{\ypart'} \weight{\ypart'}{\system(\x)} = 1$) for all systems $\system$ and inputs $\x$. 
We can now replace the sum over the components with its expectation,
\begin{eqnarray}
\Util(\system) & = &\E_{\Prob(\x)} \sum_{\ypart \in \system(\x)} \weight{\ypart}{\system(\x)} \, \util(\x, \ypart) \nonumber \\
               & = &\E_{\Prob(\x)} \E_{\Pr(\ypart \mid \x; \system)} [\util(\x, \ypart)]. \label{eq:utilmc}
\end{eqnarray}
This expectation can now be estimated via Monte Carlo, since we can sample from both $\Prob(\x)$ and $\Pr(\ypart | \x, \system)$ without expensive utility assessments. 

\subsection{Importance Sampling Estimators}

To obtain an unbiased sampling-based estimate of $\Util(\system)$ in Eq.~\eqref{eq:utilmc}, one could simply sample queries and documents from $\Prob(\x) \cdot \Pr(\ypart | \x, \system)$ and average the results. However, this naive strategy has two drawbacks. First, to evaluate each new system $\system$, it would sample documents from a new distribution, requiring additional expensive utility assessments. Second, there may be other sampling distributions that are statistically more efficient. 

In principle, we can use any unbiased Monte Carlo technique to overcome these two drawbacks of naive sampling, and \cite{yilmaz2008simple} have used stratified sampling. We deviate from their choice and focus on importance sampling for four reasons. First, importance sampling makes it straightforward to incorporate prior knowledge into the sampling distribution. This could be knowledge about the utility values $\util(x, y)$ or about the systems being evaluated. Second, we can obtain confidence intervals with little additional overhead. Third, importance sampling offers a natural and simple way to re-use previously collected judgments for evaluating new systems. Finally, as we will show in this paper, the importance sampling framework extends naturally to scenarios involving concurrent evaluation of multiple systems, providing closed-form solutions that are easy to use in practice. 

Central to importance sampling is the idea of defining a sampling distribution $Q(\x,\ypart)$ that focuses on the regions of the space that are most important for accurate estimates. We consider the family of sampling distributions that first draws a sample $\xsample$ of $\len$ queries from $\Prob(\x)$\footnote{One may also consider sampling queries $\x$ from some other distribution than $\Prob(\x)$, which may be beneficial if different types of queries contribute with different variability to the overall estimate \cite{Guiver2009,Robertson2012}.}, and then samples query-document pairs with replacement from this set. This two-step sampling via $\xsample$ has the advantage that the assessment overhead of understanding a query can now be amortized over multiple judgments per query. Note that 
we may repeatedly sample the same query-document pair. See Section~\ref{sec:noisy_feedback} for a discussion on whether to actually judge the same query-document pair more than once. 
%

For a sample of $n$ observations $\{ (\x_i,\ypart_i) \}_{i=1}^n$ drawn i.i.d. from a sampling distribution $Q(\x, \ypart)$ and a target distribution $\Pr(\x, \ypart | \system)=\Pr(\ypart | \x, \system)/\len$ for a given $\xsample$, the importance sampling estimator for Eq.~(\ref{eq:utilmc}) on $\xsample$ is
\begin{equation}
\Utilhat_n(\system) = \frac{1}{n} \sum_{i=1}^n \util(\x_i,\ypart_i) \frac{\Prob(\x_i, \ypart_i \mid \system)}{Q(\x_i,\ypart_i)}.
\label{eq:importance_estimator}
\end{equation}
Given any sampling distribution $Q(\x,\ypart)$, applying this importance sampling estimator amounts to the following simple procedure: 
\begin{enumerate}
\item Draw a sample $\xsample$ of $\len$ queries from $\Prob(\x)$. Given a budget of $n$ assessments, 
   \begin{compactitem}
   \item draw query/document pair $(\x_i,\ypart_i)$ from $Q(\x_i,\ypart)$,
   \item collect assessment $u(\x_i,\ypart_i)$ and record $Q(\x_i,\ypart_i)$.
   \end{compactitem}
   The result is a test collection $$
    \data = ((\x_i,\ypart_i,\util(\x_i,\ypart_i),Q(\x_i,\ypart_i))_{i=1}^\n .
   $$
\item For systems $\systems = \{\system_1, \ldots \system_k \}$, compute $\Utilhat_n(\system_j)$ according to Eq.~(\ref{eq:importance_estimator}) using $\data$.
\end{enumerate}

The following shows that this provides an unbiased estimate of $\Util(\system)$ \cite{mcbook}, if one ensures that $Q(\x, \ypart)$ has sufficient support, i.e. $\util(\x, \ypart) P(\x, \ypart | \system) \neq 0 \Rightarrow Q(\x, \ypart) > 0$.
\begin{eqnarray}
\E[\Utilhat_n(\system)] & = &  \E_{\Prob(\xsample)} \E_{Q(\x_i,\ypart_i)}\left[\frac{1}{n} \sum_{i=1}^n \util(\x_i,\ypart_i) \frac{\Prob(\x_i,\ypart_i \mid \system)}{Q(\x_i,\ypart_i)}\right] \nonumber \\
                        & = & \E_{\Prob(\xsample)} \E_{Q(\x, \ypart)} \left[\util(\x,\ypart) \frac{\Prob(\x, \ypart \mid \system)}{Q(\x, \ypart)}\right] \nonumber\\
                        & = & \E_{\Prob(\xsample)} \left[ \sum_{\x, \ypart} \util(\x,\ypart) \frac{\Prob(\x, \ypart \mid \system)}{Q(\x, \ypart)} Q(\x, \ypart) \right] \nonumber \\
                        & = & \E_{\Prob(\xsample)} \left[ \frac{1}{\len} \sum_{\x \in \xsample} \sum_{\ypart} \util(\x,\ypart) \Pr(\ypart \mid \x, \system) \right] \nonumber \\
                        & = & \E_{\Prob(\x)} \left[ \sum_{\ypart} \util(\x,\ypart) \Pr(\ypart \mid \x, \system) \right] \nonumber\\
                        & = & \E_{\Prob(\x)} \E_{\Pr(\ypart \mid \x; \system)} [\util(\x, \ypart)] \quad = \quad \Util(\system). \nonumber
\end{eqnarray}
Note that it is not necessary that any particular sample includes the assessment of all parts $\ypart$ of all system outputs $\y=S(\x)$ to provide an unbiased estimate. For example, when estimating Precision@10, we may get far fewer than 10 judgments per query and still have an unbiased estimator.

\subsection{Designing the Sampling Distribution Q} \label{sec:designq}

What remains to be addressed is the question of which sampling distribution $Q(\x,\ypart)$ to use for sampling the documents to be assessed for each query in $\xsample$. Since every $Q(\x,\ypart)$ with full support ensures unbiasedness, the key criterion for choosing $Q(\x,\ypart)$ is statistical efficiency \cite{aslam2006statistical}. 
Concretely, the variance $\bm{Var}_Q[ \Utilhat_n(\system) ]$ of the estimator $\Utilhat_n(\system)$ governs efficiency and is our measure of estimation quality,
\begin{equation*}
\Sigma(Q) = \bm{Var}_Q\left[ \Utilhat_n(\system) \right]. 
\end{equation*}
We therefore wish to pick a distribution $Q$ that minimizes $\Sigma(Q)$.
Let $z_Q$ be short-hand for $\util(\x,\ypart) \frac{\Prob(\x, \ypart | \system)}{Q(\x, \ypart)}$, then 
\begin{eqnarray*}
\Sigma(Q)  =  \frac{1}{\n} \bm{Var}_Q\left[z_Q\right] 
                                             = \frac{1}{\n} \left(\E_{Q}[z_Q^2] - (\E_{Q}[z_Q])^2\right).
\end{eqnarray*}
The first equality follows since $(\x_i,\ypart_i)$ are i.i.d. and $\Utilhat_n(\system)$ is a sample mean, and the second expands the definition of variance. Notice that our earlier proof of unbiasedness implies that $\mu=\E_{Q}[z_Q]=\sum_{\x, \ypart} \util(\x,\ypart) \Prob(\x, \ypart | \system)$ is a constant independent of $Q$.

%
A key result for importance sampling is that the following $Q^*$ is optimal for minimizing $\Sigma(Q)$ \cite{Kahn1953,aslam2006statistical}:
\begin{equation}
\label{eq:optsingle}
Q^*(\x,\ypart) = \util(\x, \ypart) \cdot \Prob(\x,\ypart \mid \system) / \mu.
\end{equation}
To see this, observe that for any other sampling distribution $Q$ different from $Q^*$,
\begin{align*}
    \bm{Var}_{Q^*}\left[z_{Q^*}\right] + \mu^2 & = \sum_{\x, \ypart} \frac{(\util(\x,\ypart) \Prob(\x, \ypart \mid \system))^2}{\util(\x, \ypart) \cdot \Prob(\x,\ypart \mid \system) / \mu}\\
    & = \mu \sum_{\x, \ypart} \util(\x,\ypart) \Prob(\x, \ypart \mid \system) \\
    & = (\E_{Q} \left[ z_Q \right])^2 \\
    & \le \E_{Q} \left[ {z_Q}^2 \right] = \bm{Var}_Q\left[z_{Q}\right] + \mu^2.
\end{align*} 
The first line is the definition of $\bm{Var}_{Q*}[z_{Q^*}^2]$, and the last line follows from Jensen's inequality.

Since we do not have access to the true $\util(\x, \ypart)$ in practice, one usually substitutes approximate utilities $\utilapx(\x, \ypart)$ based on prior side information (e.g. the Okapi BM25 score of document $y$ for query $x$) into Equation~\eqref{eq:optsingle}.
Any $\utilapx(\x, \ypart)>0$ retains unbiasedness, and the better the estimates $\utilapx(\x, \ypart)$, the better the efficiency of the estimator. 




\subsection{Practical Concerns}
Using the importance sampling framework outlined above offers straightforward solutions to a number of matters of practical interest.

{\bf Noisy Utility Assessments.} \label{sec:noisy_feedback}
In practice, there is often no consensus on what the true utility $\util(\x,\ypart)$ is, and different assessors (and users) will have different opinions. An example is crowd-sourcing, where labels get consolidated from multiple noisy judges. Note that our framework naturally lends itself to these noisy settings, if we think of individual assessments $\util(\x, \ypart | \user)$ as conditioned on the assessor $\user$ that is drawn from a distribution $\Prob(\user)$ and define 
$$v(\x, \ypart) = \E_{\Prob(\user)}[\util(\x, \ypart \mid \user)].$$
Our estimator (\ref{eq:importance_estimator}) stays the same, and remains unbiased by linearity of expectation. Also, the theoretical results for picking the optimal $Q^*$ remain essentially unaltered. 
The only thing that changes is that we replace the true $\util(\x, \ypart)$ with the true expected utility $v(\x, \ypart)$.

{\bf Reusing Existing Data.} Given that any sampling distribution $Q(\x,\ypart)$ with full support provides unbiased estimates, reusing old data $\data=((\x_i,\ypart_i,\util(\x_i,\ypart_i),Q(\x_i,\ypart_i))_{i=1}^\n$ is straightforward.
To guarantee full support over all query-document pairs, we can use a mixture sampling distribution such as
\begin{equation*}
Q(\x,\ypart) \propto \utilapx(\x, \ypart) \cdot P(\x,\ypart \mid \system)+ \epsilon,
\end{equation*}
where $\epsilon$ is a small constant added to ensure $Q$ has sufficiently heavy tails \cite{yuan2005heavy}.
A larger $\epsilon$ will make the collected samples more reusable for any new system, but sacrifices statistical efficiency for evaluating the current $\system$. 

In addition, it is easy to see that not all data used in the estimator from Eq.~\eqref{eq:importance_estimator} has to be collected using the same $Q(\x,\ypart)$ or the same sample of queries $\xsample$, and that we can combine datasets that accumulate over time. As long as we keep track of $Q$ for every $(\x_i,\ypart_i)$ that was sampled, each sample may be drawn from a different $Q_i(\x_i,\ypart_i)$, and we can use the mixture
\begin{eqnarray}
  Q(\x,\ypart) = \frac{1}{\n}\sum_{i=1}^{\n} Q_i(\x,\ypart), \label{eq:mixture}
\end{eqnarray}
in the denominator of our estimator. This is a direct application of the balance heuristic for Multiple Importance Sampling \cite{mcbook}. Furthermore, Equation~\eqref{eq:mixture} provides guidance on how to draw new samples given the distributions $Q_i(\x,\ypart)$ of the existing $\data$, if we eventually want the overall data to be close to a particularly efficient distribution $Q^*(\x,\ypart)$ for a new evaluation task.  

{\bf Quantifying Evaluation Accuracy.} \label{sec:confidence_intervals} An advantage of the sampling approach to evaluation is that it allows us to easily quantify the accuracy of the estimates on $\xsample$. For large sample sizes $n$, our estimate $\Utilhat_n(\system)$ converges in distribution to a normal distribution. 
From the central limit theorem, we then obtain 
\begin{equation}
\left[ \Utilhat_n(\system) - t_{\alpha/2} \frac{\hat{\sigma}_n(\system)}{\sqrt{n}},\Utilhat_n(\system) + t_{\alpha/2} \frac{\hat{\sigma}_n(\system)}{\sqrt{n}} \right]
\label{eq:CLT}
\end{equation}
as the approximate $1-\alpha$ confidence interval. For example, a 95\% confidence interval would be $\Utilhat_n(\system) \pm 1.96 \frac{\hat{\sigma}_n(\system)}{\sqrt{n}}.$ 
We can estimate $\hat{\sigma}_n(\system)$ from the same samples $\{ (\x_i,\ypart_i) \}_{i=1}^n$ that we used to compute $\Utilhat_n(\system)$ as follows:
$$\hat{\sigma}_n(\system) = \sqrt{\frac{1}{n-1} \sum_{i=1}^{n} \left(\frac{\util(\x_i, \ypart_i)\Prob(\x_i,\ypart_i \mid \system)}{Q(\x_i,\ypart_i)} - \Utilhat_n(\system)\right)^2}.$$
\noindent
Note that the rate of convergence of $\Utilhat_n(S)$ to its true value depends on the skewness 
of the distribution of $z_Q$, and we will empirically evaluate the quality of the confidence intervals in Section~\ref{sec:expt_singlesystem}.

\section{Comparing Two Systems} \label{sec:two_comp}


Up until now this paper has only considered the problem of estimating the performance of one system in isolation. In practice, however, we are typically much more interested in the relative performance of multiple systems. In the case of two systems $\system$ and $\system'$, we may be interested in measuring how much they differ in performance $\Diff(\system,\system') = \Util(\system) - \Util(\system')$. 
To this effect, we consider the estimator
\begin{eqnarray}
\Diffhat_n(\system,\system') = \frac{1}{n} \sum_{i=1}^n \util(\x_i,\ypart_i) \frac{\Prob(\x_i, \ypart_i \mid \system) - \Prob(\x_i, \ypart_i \mid \system')}{Q(\x_i,\ypart_i)}.
\label{eq:importance_estimator_diff}
\end{eqnarray}
Again, this estimator is unbiased, and it can be computed using data $\data=((\x_i,\ypart_i,\util(\x_i,\ypart_i),Q(\x_i,\ypart_i))_{i=1}^\n$ sampled from any $Q$ with sufficient support. 
But what does the most efficient sampling distribution $Q^*$ look like? Like in the single system case, the sample efficiency of the estimator is governed by its variance,



\begin{equation}
\Sigma(Q) = \bm{Var}_Q\left[ \Diffhat_\n(\system,\system') \right]. \label{eq:sigma_q_pair}
\end{equation}
Analogous to $\Utilhat_\n(\system)$, $\Diffhat_\n(\system,\system')$ is the average of $n$ i.i.d. random variables $z_Q = \util(\x,\ypart) \frac{\Prob(\x, \ypart | \system) - \Prob(\x, \ypart | \system')}{Q(\x, \ypart)}$.
So, $\Sigma(Q)$ is proportional to $\bm{Var}_Q\left[ z_Q \right]$ 
and the only term that depends on $Q$ is $\E_Q \left[ (z_Q)^2 \right]$.
Following a similar argument as before, the optimal sampling distribution is
\begin{equation}
\label{eq:qstar_pair}
Q^*(\x, \ypart) \propto \util(\x, \ypart)\cdot \left| P(\x, \ypart | \system) - P(\x, \ypart \mid \system') \right|.    
\end{equation}
$\E_{Q^*} [ (z_{Q^*})^2] = \E_{Q} [| z_Q | ]^2 \le \E_{Q} [ (z_Q)^2]$ by Jensen's inequality, so $Q^*$ from Equation~\eqref{eq:qstar_pair} minimizes $\Sigma(Q)$.
Unfortunately, we again cannot compute $Q^*$ because it needs $\util(\x,\ypart)$, but we can substitute an approximate $\utilapx(\x,\ypart)$ as before.

Note that this $Q^*$ is very intuitive -- items that have similar weights $P(\x, \ypart | \cdot)$ in both systems will get sampled with a low probability, since they have negligible effect on the performance difference. 
This $Q^*$ is different from the heuristic $Q$ previously used in multi-system evaluation \cite{yilmaz2008simple}, where $Q$ is simply the average of $P(\x, \ypart | \system)$ and $P(\x, \ypart | \system')$. In particular, this heuristic $Q$ fails to recognize that documents at identical positions in both rankings contribute no information about the performance difference.
In Section~\ref{sec:expt_2system} we empirically compare against this heuristic $Q$.


\section{Comparing Multiple Systems to a Baseline} \label{sec:k_baseline}

We now consider another evaluation use case that is frequently encountered in practice. We have a current production system $\system'$ and several new candidate systems $\systems = \{\system_1, \ldots \system_k \}$. The goal of evaluation is to estimate by how much each candidate improves (or not) over the baseline $\system'$.

We can formulate this goal in terms of $k$ comparative evaluation problems $\Diff(\system_i,\system') = \Util(\system_j) - \Util(\system')$, and we want reliable estimates for all performance differences $\Diff(\system_1,\system')$, ..., $\Diff(\system_k,\system')$.
We can use the estimator $\Diffhat_\n(\system_j,\system')$ from Equation~\eqref{eq:importance_estimator_diff} for each $\Diff(\system_j,\system')$, and the procedure for computing it is identical to Section~\ref{sec:two_comp}.
This is one of the strengths of the sampling approach: once a sampling distribution $Q$ is designed, multiple systems and multiple differences can be concurrently evaluated from one batch of judgments using a unified, unbiased procedure. 

But what is the optimal $Q^*$ for this new use case? 
Since we are now considering $k$ estimates in parallel, we consider the sum of estimator variances as our measure of estimation quality to optimize: 
\begin{equation}
\Sigma(Q) = \sum_{j=1}^k \bm{Var}_Q\left[ \Diffhat_\n(\system_j,\system') \right]. \label{eq:sigma_q}
\end{equation}
We will show that the distribution minimizing $\Sigma(Q)$ is
\begin{equation}
\label{eq:qstar_baseline}
Q^*(\x, \ypart) \propto \util(\x, \ypart)\cdot \sqrt{\sum_{j=1}^k (P(\x, \ypart \mid \system_j) - P(\x, \ypart \mid \system'))^2}.    
\end{equation}
To see this, collect the terms of $\Sigma(Q)$ that depend on $Q$ (up to a scaling constant $n$) and denote them as,
\begin{equation*}
T(Q) = \E_Q \left[ \left(\frac{\util(\x,\ypart)}{Q(\x,\ypart)}\right)^2 \sum_{j=1}^k (P(\x,\ypart \mid \system_j) - P(\x,\ypart \mid \system'))^2 \right].
\end{equation*}
We can then show that
\begin{align*}
T(Q^*) & = \left[ \sum_{\x,\ypart} \util(\x, \ypart) \sqrt{\sum_{j=1}^k (P(\x, \ypart \mid \system_j) - P(\x, \ypart \mid \system'))^2} \right]^2 \\
&= \E_{Q} \left[ \frac{\util(\x, \ypart)}{Q(\x,\ypart)} \sqrt{\sum_{j=1}^k (P(\x, \ypart \mid \system_j) - P(\x, \ypart \mid \system'))^2}\right]^2 \\
& \le \E_{Q} \left[ \left(\!\frac{\util(\x, \ypart)}{Q(\x,\ypart)}\!\right)^{\!2} \sum_{j=1}^k (P(\x, \ypart \mid \system_j) - P(\x, \ypart \mid \system'))^2 \right] \\
& = T(Q).
\end{align*}

As before we use an approximate $\utilapx(\x, \ypart)$ for a computable sampling distribution in our experiments.

Our results here complement recent work in multidimensional importance sampling \cite{Zhao2015} which studies multidimensional $\util(\x,\ypart)$ and a single target distribution $P(\x,\ypart | \system)$.
The choice of sum-of-variances as the $\Sigma(Q)$ objective yields a simple closed-form optimal sampling distribution $Q^*$ for this evaluation scenario. Several other $\Sigma(Q)$ are possible (for instance, the maximum variance $\max_j \bm{Var}_Q[ \Diffhat_\n(\system_j,\system') ]$),
however closed form $Q^*$ that optimize these may not exist.
We defer further study of different objectives characterizing estimation quality in these scenarios to future work.

\section{Ranking Multiple Systems} \label{sec:k_relative}

Finally, we consider the use-case of ranking a collection of systems $\systems = \{\system_1, \ldots \system_k \}$ in order of their performance $\{ \Util(\system_1), \ldots, \Util(\system_k) \}$. A first thought may be to estimate each $\Util(\system_j)$ directly, which we call the absolute evaluation strategy. However, we can often do much better using comparative evaluations, since accurate ranking merely requires that we can estimate $\{ \Util(\system_1)+\delta, \ldots, \Util(\system_k)+\delta \}$ up to some arbitrary constant $\delta$. 

\begin{sloppypar}
Why should this comparative problem be more accurate than absolute evaluation? Imagine three systems, two of which are poor ($\Util(\system_1), \Util(\system_2) \simeq 0$) while the third is good ($\Util(\system_3) \gg 0$). 
Consider the case where there are a lot of ``easy'' documents that all three system rank correctly, and that the large difference between $\system_3$ and systems $\system_1$/$\system_2$ is due to a few ``hard'' documents where $\system_3$ is superior. Furthermore, $\system_1$ and $\system_2$ may be small variants of the same system that produce almost identical rankings. Designing $Q^*$ to optimize for the sum of absolute variances $\sum_{j=1}^3 \bm{Var}_Q[ \Utilhat_n(\system_j) ]$ would be ignorant to all this structure.
\end{sloppypar}


To design a more informed estimator and sampler for ranking, we propose to merely estimate each system's performance relative to some baseline system $\tau$,
\begin{eqnarray}
    \Diff(\system_1,\tau), \ldots \Diff(\system_k,\tau),
\end{eqnarray}
and then rank the systems using the estimator $\Diffhat(\system_i,\tau)$ from Equation~\eqref{eq:importance_estimator_diff}. We could then use the optimal sampling distribution $Q^*$ from Section~\ref{sec:k_baseline} to sample query-document pairs and collect judgments. 

What remains to be shown is how to construct the optimal baseline system $\tau$. Using $$\Sigma(\tau)=\sum_{j=1}^k \bm{Var}_{Q^*}\left[ \Diffhat_\n(\system_j,\tau) \right]$$ as our measure of estimator quality analogous to the previous sections, we prove that the optimal $\tau$ corresponds to
$$
    P(\x, \ypart | \tau) = \frac{1}{k} \sum_{j=1}^k P(\x, \ypart \mid \system_j)
$$
for each query-document pair $(\x, \ypart)$.
\begin{proof}
From the results of Section~\ref{sec:k_baseline}, we know that for any system $\tau$, the optimal sampling distribution $Q^*(x,y) \propto \util(\x, \ypart)\cdot \sqrt{\sum_{j=1}^k (P(\x, \ypart | \system_j) - P(\x, \ypart | \tau))^2}$. If we plug in this $Q^*$ into $\Sigma(\tau)$ and simplify (note that all terms of the variance depend on $\tau$),
\begin{eqnarray*}
    \Sigma(\tau) &=& \left[ \sum_{\x, \ypart} \util(x,y)\sqrt{\sum_{j=1}^k (P(\x, \ypart \mid \system_j) - P(\x, \ypart \mid \tau))^2} \right]^2 \\
    & & - \sum_{j=1}^k \left[ \sum_{\x, \ypart} \util(\x,\ypart) \{ P(\x, \ypart \mid \system_j) - P(\x, \ypart \mid \tau) \} \right]^2.
\end{eqnarray*}

Let each $P(\x, \ypart | \tau)$ be a variable $\tau_{\x,\ypart}$ which we minimize over. $\Sigma(\tau)$ is convex in each $\tau_{\x,\ypart}$.
A minimum requires that $\partial \Sigma(\tau) / \partial \tau_{\x,\ypart} = 0$, subject to $\tau_{\x,\ypart} \ge 0$ for all $(\x,\ypart)$ and $\sum_{\x, \ypart} \tau_{\x,\ypart} = 1$.
To begin, $\partial \Sigma(\tau) / \partial \tau_{\x_i,\ypart_i} = 0$ yields that $\forall (\x_i, \ypart_i)$:
\begin{align*}
    \sum_{\x,\ypart} \!\util(\!\x,\! \ypart\!) \!\!\left[ \sum_{j=1}^k\!\! P(\!\x,\! \ypart | \system_j\!) \!-\! \tau_{\x,\ypart} \!\right] \!+ \alpha \! \sum_{j=1}^k\! P(\!\x_i,\! \ypart_i | \system_j\!) \!-\! \tau_{\x_i,\ypart_i} \!=\! 0\\ 
    \text{where} \quad
    \alpha = \frac{\sum_{\x,\ypart} \util(\x,\ypart) \sqrt{\sum_{j=1}^k (P(\x, \ypart \mid \system_j) - \tau_{\x,\ypart})^2}}{\sqrt{\sum_{j=1}^k (P(\x_i, \ypart_i \mid \system_j) - \tau_{\x_i,\ypart_i})^2}}.
\end{align*}
Observe that $\tau_{\x_i,\ypart_i} = \frac{1}{k}\sum_{j=1}^k P(\x_i,\ypart_i | \system_j)$ satisfies all the equations simultaneously, which completes the proof. 
\end{proof}

Putting everything together, the optimal sampling distribution is
\begin{equation}
\label{eq:qstar_ranking}
\!\!\!Q^*\!(\x, \ypart) \propto \util(\x, \ypart) \sqrt{\!\sum_{j=1}^k \!\!\left(\!\!P(\x, \ypart | \system_j) \!-\! \frac{1}{k} \!\sum_{i=1}^k \!P(\x, \ypart | \system_i)\!\!\right)^{\!\!2}},    
\end{equation}
and we can again use an approximate $\utilapx(\x, \ypart)$ in practice.

\section{Experiments}
\label{sec:expt}
The following experiments evaluate to what extent the theoretical contributions developed in this paper impact evaluation accuracy empirically. We compare sampling-based approaches only in this section, recognizing that deterministic approaches do not deliver the same guarantees as laid out in the introduction.  To study effects in isolation, we first explore our estimator and sampling distribution design principles on the problem of single-system evaluation. We then in turn consider the three comparative multi-system evaluation problems. 


\begin{table}[t]
  \centering
      \setlength{\tabcolsep}{3pt} 
    \begin{tabular}{rrrrrrr}
    \toprule
      & \parbox{1cm}{\centering True $\Util(\system)$} & \parbox{1cm}{\centering Shallow Pool} & \parbox{2cm}{\centering Deep \linebreak[4] Pool} & \parbox{2cm}{\centering Sampling \linebreak[4] $\Utilhat(\system)$} \\
      \midrule
        OPT   & 284.40 & 16.98 & 284.48 $\pm$ 1.69 & 284.54 $\pm$ 1.22 \\
        REV-75 & 277.63 & 10.00 & 277.79 $\pm$ 1.75 & 277.58 $\pm$ 1.07 \\
        REV-150 & 271.32 & 8.51 & 271.21 $\pm$ 1.59 & 271.18 $\pm$ 0.97 \\
        SHIFT-5 & 274.94 & 4.72 & 275.03 $\pm$ 1.69 & 274.73 $\pm$ 1.10 \\
        SHIFT-7 & 269.87 & 0.00 & 269.90 $\pm$ 1.75 & 269.98 $\pm$ 1.12 \\
        \midrule
        mds08a3 & 6.96  & 1.93 & 7.32 $\pm$ 5.05 & 6.86 $\pm$ 0.76 \\
        nttd8ale & 9.01  & 2.41 & 8.79 $\pm$ 5.30 & 8.97 $\pm$ 0.87 \\
        weaver2 & 7.48  & 1.94  & 8.40 $\pm$ 6.59 & 7.53 $\pm$ 0.83   \\
    \bottomrule
    \end{tabular}%
  \caption{Mean and standard deviation of DCG estimates across 100 trials for all systems in SYNTH (top) and three randomly chosen systems in TREC (bottom). The first column shows the true $\Util(\system)$ that ShallowPool, DeepPool, and $\Utilhat(\system)$ from Equation~\eqref{eq:importance_estimator} (with approximate $Q^*$ sampling) aim to estimate.}
  \label{tbl:det_vs_sampling}%
\end{table}%

\subsection{Datasets and Experiment Setup} \label{sec:expsetup}
We use two datasets for our experiments, \emph{SYNTH} and \emph{TREC}, which give us different levels of experimental control and different application scenarios. The SYNTH dataset is designed to resemble judgments as they occur in a recommender system. To create SYNTH, we generated a sample $\xsample$ of 6000 users (i.e. queries) and 2000 items (i.e. documents). The ground truth judgments $\util(x, y) \in \{0,...4\}$ for each user/item pair were drawn from a categorical distribution whose parameters were drawn from a Dirichlet distribution with hyper-parameters $\alpha = (.54, .25, .175, .03, .005)$, so as to give high ratings a low probability.
Based on this data, we created ranking systems $\system_j$ of different quality in the following way. 
$\system_{OPT}$ denotes the perfect ranking system where each user ranking $\y$ is sorted according to the true relevances $\util(x, y)$. The $\system_{SHIFT-m}$ ranking system shifts all rankings of $S_{OPT}$ down by $m$ entries; elements that get shifted beyond the last position get re-introduced at the first. $\system_{REV-m}$ reverses the order of the top $m$ elements in $S_{OPT}$. The collection of systems was $\systems = \{\system_{OPT}, \system_{REV-75}, \system_{REV-150}, \system_{SHIFT-5}, \system_{SHIFT-7}\}$. As evaluation measure, we use DCG@2000.

The TREC dataset contains binary relevance judgments and mimics a typical information retrieval scenario. To generate the dataset, we start with 
the TREC-8 ad hoc track \cite{trec8}, which comprises 149 systems that submitted rankings $\y$ of size 1000 as response to 50 queries $\xsample$. To eliminate unwanted biases due to unjudged documents which could confound our empirical evaluation, we only consider a random subset of 20 systems for which we ensured that all the top 100 documents of each of the 50 queries were fully judged. Correspondingly, we truncated all rankings to length 100 and evaluate in terms of DCG@100. 

{\bf Approximate Utilities.} If not noted otherwise, we use approximate utilities $\utilapx_\system(\x, \ypart)$ when designing the sampling distribution in all of our experiments. We use the following simple heuristic to define $\utilapx_\system(\x, \ypart)$ and we will evaluate empirically to what degree this can be improved.
For SYNTH and any given system $\system$, we use $\utilapx_\system(\x, \ypart) = 4\cdot (1-\frac{\rank(\system(\x), \ypart)}{2000})$ -- reflecting the fact that relevances ranged between 0 and 4. We have an analogously rank-decreasing function $\utilapx_\system(\x, \ypart) = \frac{16}{\rank(\system(\x), \ypart) + 34}$ for TREC. 
To define approximate utilities for a set of systems $\systems=\{\system_1, \ldots, \system_k\}$, we simply average the $\utilapx_{\system_j}(\x, \ypart)$.

{\bf Comparing systems.} When performing comparative evaluations in Sections~\ref{sec:expt_2system} to \ref{sec:expt_ranking}, we focus on the most difficult comparisons in the following way. We rank all systems by their true performance $\Util(S)$ on our ground truth set, and then compare adjacent systems in Section~\ref{sec:expt_2system} and a sliding window of five systems in Sections~\ref{sec:expt_baseline} and \ref{sec:expt_ranking}. In Section~\ref{sec:expt_baseline}, the middle system is used as the baseline $\system'$. Each experiment is replicated $100$ times.

\begin{table}[t]
  \centering
    \begin{tabular}{rrrr}
    \toprule
          &  $Q^*$ & $Q^{P}$ & $Q^{unif}$ \\
    \midrule
    OPT        &  1.22 & 1.98 & 3.05 \\
    REV-75 &  1.07 & 1.45 & 2.64 \\
    REV-150 & 0.97 & 1.33 & 2.20 \\
    SHIFT-5 &     1.10 & 1.67 & 2.63 \\
    SHIFT-7 &     1.12 & 1.45 & 2.45 \\
    \midrule
    mds08a3 &     0.76 & 0.84 & 0.97 \\
    nttd8ale &    0.87 & 0.98 & 1.18 \\
    weaver2 &     0.83 & 0.92 & 1.05 \\
    \bottomrule
    \end{tabular}%
  \caption{Standard deviation of $\Utilhat(\system)$ for DCG@k estimates under different sampling methods across $100$ trials for all systems in SYNTH (top) and three randomly chosen systems in TREC (bottom).}
  \label{tbl:prior_knowledge}%
\end{table}%

\subsection{Empirical Verification of Design Principles} \label{sec:expt_singlesystem}

We first evaluate how the design of the sampling distribution $Q(x, y)$ affects the efficiency of the $\Utilhat(\x,\ypart)$ estimator in Equation~\eqref{eq:importance_estimator}. To study the effect in isolation and in comparison to conventional pooling methods, we first focus on the single-system evaluation problem.

{\bf Comparison to Deterministic Pooling.}
To ground our experiments with respect to conventional practice, we start by comparing the sampling approach to deterministic pooling \cite{SparckJones1975} historically used in TREC. 
The depth of the pool $b$, i.e., the number of top-$b$ documents being judged, vs the number of queries $\len$ is the key design choice when pooling under a judgment budget. We use two pooling methods that choose this trade-off differently. Given a budget of $n$ assessments, the \textit{ShallowPool} methods gets judgments for the top $b' = \frac{n}{\len}$ items of each query and computes the average performance based on these judgments. The \textit{DeepPool} method 
randomly selects $l' = \frac{n}{b}$ queries and judges all pooled documents for those queries.

Table~\ref{tbl:det_vs_sampling} compares estimated DCG@k 
for the deterministic baselines ShallowPool and DeepPool with the sampling estimates $\Utilhat(\system)$ from Eq.~(\ref{eq:optsingle}) with a total budget of $n=5\cdot |\xsample|$ assessments. We repeated each experiment $100$ times, and average the estimates across all 100 runs. The error intervals correspond to the empirical standard deviations of the estimates. Note that ShallowPool has standard deviation of zero, since it is deterministic given $\xsample$.

Unsurprisingly, Table~\ref{tbl:det_vs_sampling} shows that ShallowPool is heavily biased by the missing judgments and systematically underestimates the true $\Util(\system)$ given in the first column. The bias of ShallowPool is so strong that even its average estimates do not reflect the correct ordering of the ranking functions 
on SYNTH. 

Table~\ref{tbl:det_vs_sampling} shows that our $\Utilhat(\system)$ estimator is indeed unbiased. DeepPool is also unbiased, since it has a non-zero probability of eliciting judments for any returned document.
However, $\Utilhat(\system)$ has substantially lower standard deviation especially on TREC. This is not surprising, since DeepPool suffers from between-query variability when sampling only a few queries from $\xsample$. In fact, the standard deviations of DeepPool are much larger than the estimated differences on TREC, indicating that DeepPool cannot reliably distinguish the ranking performance of the systems.


{\bf Impact of Prior Knowledge.} 
We now study the impact of the sampling distribution on sampling efficiency.
Table~\ref{tbl:prior_knowledge} compares three sampling approaches: $Q^*$ with approximate utilities as in the previous experiment, $Q^{P}$ with $\utilapx(x,y) = 1$, and uniform sampling $Q^{unif}$.

We see in Table~\ref{tbl:prior_knowledge} that the standard deviation increases as we transition from $Q^*$ to $Q^P$, and it increases even further if we sample uniformly in $Q^{unif}$. Note that an estimator with half the standard deviation requires only roughly half the sample size to reach the same level of estimation accuracy.
We conclude that the choice of $Q$ is of great practical significance. This is in line with importance sampling theory that indicates that good estimates of $\utilapx$ can substantially improve estimation quality. Given these findings, all following experiments will be based on the $Q^*$ sampler with approximate utilities.


{\bf Confidence Intervals.}
Another feature of the sampling-based approach is that one can quantify its accuracy using confidence intervals. But are the approximate confidence intervals proposed in Section~\ref{sec:confidence_intervals} accurate in practice?
For each of the 100 runs from Table~\ref{tbl:det_vs_sampling}, we computed 95\% confidence intervals according to Equation \eqref{eq:CLT}. We also computed the coverage probability $\hat{P}_{\textrm{cov}}$ of the confidence intervals, defined as the number of times the true value was within the confidence interval. As we can see from Table~\ref{tbl:confidence_intervals}, the coverage probabilities of the approximate intervals are usually very close to 95\% as desired. We can also verify that the empirical standard deviation of $\Utilhat(\system)$ is roughly $1/1.96$ of the average confidence intervals.

\begin{table}[t]
  \centering
\begin{tabular}{rrrr}
\toprule
      & $\Utilhat(\system)$ & ConfInt & $\hat{P}_{\textrm{cov}}$  \\
\midrule
OPT   & 284.54 $\pm$ 1.22 & $\pm$ 2.21 & 0.92   \\
REV-75 & 277.58 $\pm$ 1.07 & $\pm$ 2.10 & 0.96   \\
REV-150 & 271.18 $\pm$ 0.97 & $\pm$ 2.05 & 0.95   \\
SHIFT-5 & 274.73 $\pm$ 1.10 & $\pm$ 2.18 & 0.93   \\
SHIFT-7 & 269.98 $\pm$ 1.12 & $\pm$ 2.18 & 0.94   \\
\midrule
mds08a3 & 6.86 $\pm$ 0.76 & $\pm$ 1.50 & 0.95   \\
nttd8ale & 8.97 $\pm$ 0.87 & $\pm$ 1.65 & 0.94   \\
weaver2 & 7.53 $\pm$ 0.83   & $\pm$ 1.58 & 0.95   \\
\bottomrule
\end{tabular}%
\caption{Average confidence intervals and coverage probabilities for the $\Utilhat(\system)$ estimator using $Q^*$.}
\label{tbl:confidence_intervals}
\end{table}

\subsection{Is pairwise comparative evaluation more efficient than individual evaluation?}
\label{sec:expt_2system}

We now evaluate how far comparative estimation can improve upon individual system evaluation.
Prior work \cite{yilmaz2008simple} uses $Q(\x,\ypart) \propto \utilapx(\x,\ypart) (\Prob(\x,\ypart|\system)+\Prob(\x,\ypart|\system'))/2$ 
which we compare to the pairwise $Q^*(\x,\ypart)$ with approximate utilites from Equation~\eqref{eq:qstar_pair}.  

Table~\ref{tbl:pairwise_comparison} shows the estimator variance $\Sigma(Q)$ as defined in Equation~\eqref{eq:sigma_q_pair} for both sampling methods (up to scaling by the sample size $\n$). On both datasets, the pairwise $Q^*$ substantially improves estimator accuracy over Naive sampling. Note that reducing estimator variance by a quarter corresponds to halving the sample size needed to get a particular level of estimation accuracy. The table also contains the variance $\Sigma(Q)$ when using the true utilities $\util^*(\x,\ypart)$ for the design of the sampling distribution instead of the approximate utilities $\utilapx(\x,\ypart)$. Comparing the variance gains between Naive and $Q^*$ to the gains we could get by moving from $\utilapx(\x,\ypart)$ to $\util^*(\x,\ypart)$, we see that the pairwise $Q^*$ has improved sample efficiency far more than we could still hope to improve by finding a better $\utilapx(\x,\ypart)$ for these datasets. Note that all numbers in Table~\ref{tbl:pairwise_comparison} were computed analytically, so there are no errorbars.

\begin{figure}[t]
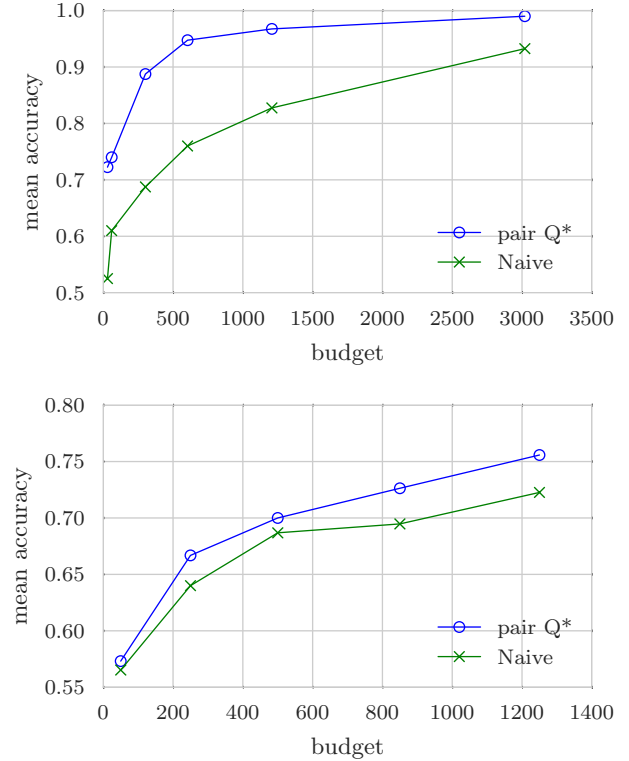

    \hspace*{-0.3cm}\input{figures/pairwise_synth.pgf} \\[0cm]
    \hspace*{-0.3cm}\input{figures/pairwise_trec.pgf}
    \vspace{-1.5em}
\caption{Mean pairwise accuracy on the SYNTH dataset (top) and the TREC dataset (bottom).} 
\label{fig:pairwise_acc}
\end{figure}

While Table~\ref{tbl:pairwise_comparison} measures estimator accuracy in terms of mean-squared-error, we are often merely interested in the binary decision of which system is better, i.e. $\sign(\Diff(\system,\system'))$. Figure~\ref{fig:pairwise_acc} compares the accuracy with which the sign of our estimate $\sign(\Diffhat(\system,\system'))$ predicts the true order of the systems on our query sample $\xsample$ correctly, $Acc = \Ind[\Diff(\system,\system') \cdot \Diffhat(\system,\system') > 0]$. On both datasets, the pairwise $Q^*$ sampling outperforms Naive. The magnitude of the gain is largest on the SYNTH dataset, where the rankings produced by different systems are more similar to each other than for the TREC systems.

\begin{table}[t!]
  \centering
\begin{tabular}{rrrrrr}
\toprule
      & \multicolumn{2}{c}{SYNTH} & & \multicolumn{2}{c}{TREC} \\
       \cline{2-3}  \cline{5-6}
      & $\utilapx(x, y)$ & $\utilopt(x, y)$& \phantom{$I^{2^{2}}$}      & $\utilapx(x, y)$ & $\utilopt(x, y)$ \\
      \midrule
Naive & 2.15  & 1.13 &  & 6.60 & 4.65 \\
pair $Q^*$ & 0.24  & 0.21 &  & 1.45  & 0.77 \\
\bottomrule
\end{tabular}%
\caption{Variance $\Sigma(Q) \cdot \n$ defined in Equation~\eqref{eq:sigma_q_pair} for the pairwise comparison problem. The table compares naive averaging with our optimal $Q^*$ from Equation~\eqref{eq:qstar_pair} under perfect and approximate knowledge of utilities.}
\label{tbl:pairwise_comparison}
\end{table}%

\subsection{Is comparative evaluation against a baseline better than individual evaluation?} \label{sec:expt_baseline}

\begin{figure}[t]
  \centering
    \hspace*{-0.3cm}\input{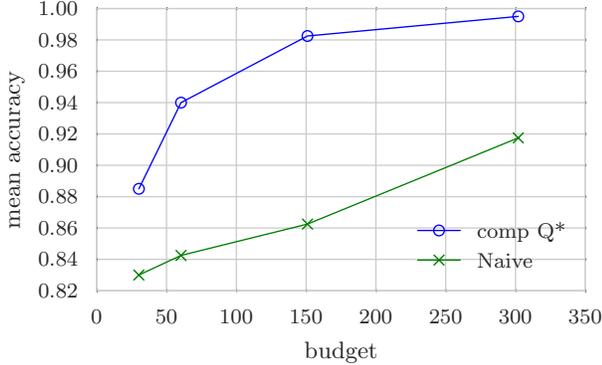} \\[0cm]
    \hspace*{-0.3cm}\input{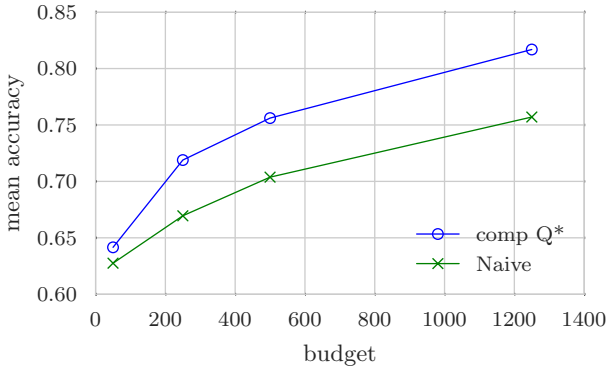}
    \vspace{-1.5em}
\caption{Average accuracy when comparing four new systems against a baseline system on the SYNTH dataset (top) and the TREC dataset (bottom).}
\label{fig:krelb_acc}
\end{figure}

We now evaluate the statistical efficiency of our method for the problem of comparing multiple systems against a baseline. Table~\ref{fig:krelb_acc} shows the aggregate variance $\Sigma(Q)$ according to Eq.~\eqref{eq:sigma_q} 
computed analytically for both the SYNTH and the TREC datasets. The table shows that our $Q^*$ for comparative evaluation as defined in Eq.~\eqref{eq:qstar_baseline} substantially outperforms the naive heuristic of sampling according to $Q(\x,\ypart) \propto \utilapx(\x,\ypart) \sum_j \Prob(\x,\ypart|\system_j)$ (which originates from Multiple Importance Sampling and Mixture Importance Sampling \cite{mcbook}). The gain is particularly large for the realistic case of approximate utilities.

Figure~\ref{fig:krelb_acc} evaluates the accuracy with which the estimated $\Ind[\Diffhat(\system_j,\system') \ge 0 ]$ predicts the true $\Ind[\Diff(\system_j,\system') \ge 0 ]$. Mean accuracy is defined as $Acc = \Ind[\Diff(\system_j,\system') \cdot \Diffhat(\system_j,\system') > 0]$ over all comparisons using the sliding window approach described in Section~\ref{sec:expsetup}. Again, we see large gains in efficiency from using our $Q^*$ over naive averaging -- both samplers using approximate utilites -- reducing the required sample size for a desired accuracy by a half or more on both datasets.

\begin{table}[t]
  \centering
\begin{tabular}{rrrrrr}
\toprule
      & \multicolumn{2}{c}{SYNTH} & & \multicolumn{2}{c}{TREC} \\
       \cline{2-3}  \cline{5-6}
      & $\utilapx(x, y)$ & $\utilopt(x, y)$& \phantom{$I^{2^{2}}$}      & $\utilapx(x, y)$ & $\utilopt(x, y)$ \\
      \midrule
Naive & 1.31  & 0.70 &  & 15.08 & 1.77 \\
comp $Q^*$ & 0.18  & 0.15 &  & 6.82  & 1.28 \\
\bottomrule
\end{tabular}%
\caption{Aggregate variance $\Sigma(Q) \cdot \n$ defined in Equation~\eqref{eq:sigma_q} for the problem of comparing against a baseline. The table compares naive averaging with our optimal $Q^*$ from Equation~\eqref{eq:qstar_baseline} under perfect and approximate knowledge of utilities.}
\label{tbl:rel_perf}
\end{table}%

\begin{table}[t]
  \centering
\begin{tabular}{rrrrrr}
\toprule
      & \multicolumn{2}{c}{SYNTH} & & \multicolumn{2}{c}{TREC} \\
       \cline{2-3}  \cline{5-6}
      & $\utilapx(x, y)$ & $\utilopt(x, y)$& \phantom{$I^{2^{2}}$}      & $\utilapx(x, y)$ & $\utilopt(x, y)$ \\
      \midrule
Naive & 0.86  & 0.46  &       & 38.64 & 1.79 \\
rank $Q^*$ & 0.11  & 0.09  &       & 12.40 & 1.12 \\
\bottomrule
\end{tabular}%
\caption{Aggregate variance $\Sigma(Q) \cdot \n$ defined in Equation~\eqref{eq:sigma_q} for the ranking problem. The table compares naive averaging with our optimal $Q^*$ from Equation~\eqref{eq:qstar_ranking} under perfect and approximate knowledge of utilities.}
\label{tbl:rel_perf_variance}%
\end{table}%

\begin{figure}[t]
  \centering
    \hspace*{-0.3cm}\input{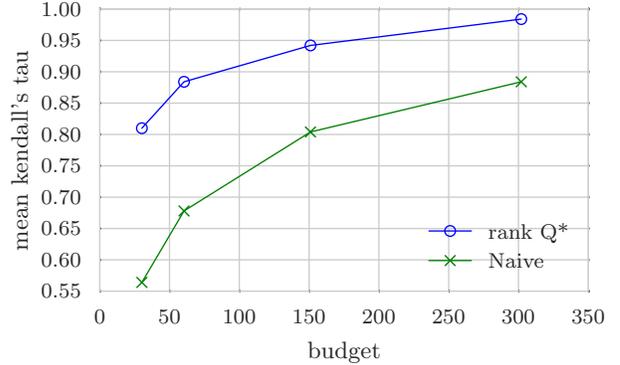} \\[0cm] 
    \hspace*{-0.3cm}\input{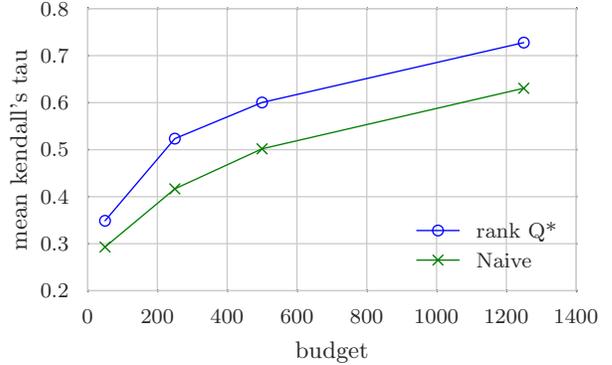}
    \vspace{-1.5em}
\caption{Kendall's tau for one set of 5 systems on SYNTH (top) and 16 sets of 5 systems on TREC (bottom).}
\label{fig:krel_acc}
\end{figure}

\subsection{Is comparative ranking more efficient than individual evaluation?} \label{sec:expt_ranking}

Finally, we turn to the problem of estimating the relative performance of $k$ systems. Following the pattern from the previous two subsections, we first compare our $Q^*$ from Eq.~\eqref{eq:qstar_ranking} with the naive averaging sampler 
in terms of their aggregate variance $\Sigma(Q)$ as defined in Eq.~\eqref{eq:sigma_q}. Sets of systems to compare were chosen using the sliding-window approach described in Section~\ref{sec:expsetup}. Again, the gains in statistical efficiency are especially large for the practical case, where the utilities are approximated.

Figure~\ref{fig:krel_acc} evaluates to what extent this gain in mean squared estimation error translates into an improved accuracy for system ranking. In particular, we measure Kendall's tau when ranking by $\Diffhat(\system_j,\tau)$, which is identical to ranking by $\Utilhat(\system_j)$. Analogous to the previous experiments, we see that our $Q^*$ sampler for ranking is again substantially more accurate than naive sampling.



\section{Conclusions and Future Work}

We developed a general and practical framework for evaluating ranking systems, making explicit the tight connections to Monte-Carlo estimation. This formal framework brings improved clarity and generality to the problem of sampling-based evaluation, including the design of estimators for new performance measures, conditions for unbiasedness, the reuse of data, and the design of sampling distributions. In particular, we focused on the question of how to design estimators and sampling distributions for comparative system evaluations, deriving variance-optimizing strategies for pairwise evaluation, comparing $k$-systems against a baseline, and ranking $k$ systems. Empirical results show that these evaluation strategies lead to substantial improvements over previously used heuristics. 

There are many directions for future work. First, much of the methodology should be applicable in other complex evaluation tasks as well, e.g., in natural language processing. Second, it would be interesting to see how to better approximate $\util(x, y)$ by learning from previously collected judgments. This could potentially also be done in an adaptive fashion -- similar to adaptive importance sampling. 
Third, there are various other evaluation questions involving $k$ systems, e.g., (adaptive) sampling so as to maximize the probability of finding the best out of $k$ systems.

\iftoggle{ARXIV}{
\section{Acknowledgments}
This work was supported in part through NSF Awards IIS-1247637, IIS-1217686, IIS-1513692, and a gift from Bloomberg.
}

\bibliographystyle{abbrv}
\bibliography{references}  
\end{document}